\documentclass[11pt,twoside]{article}

\usepackage{amsmath,epsf,amssymb,cite}

\newtheorem{theorem}{Theorem}
\newtheorem{prop}{Proposition}

\newif\iffigs\figstrue

\DeclareFontFamily{U}{rsf}{}
\DeclareFontShape{U}{rsf}{m}{n}{
  <5> <6> rsfs5 <7> <8> <9> rsfs7 <10-> rsfs10}{}
\DeclareMathAlphabet\Scr{U}{rsf}{m}{n}

\def\pplogo{\vbox{\kern-\headheight\kern -29pt
\halign{##&##\hfil\cr&{
\ppnumber}\cr\rule{0pt}{2.5ex}&\ppdate\cr}
}}
\makeatletter
\def\ps@firstpage{\ps@empty \def\@oddhead{\hss\pplogo}%
  \let\@evenhead\@oddhead 
}
\def\maketitle{\par
 \begingroup
 \def\thefootnote{\fnsymbol{footnote}}
 \def\@makefnmark{\hbox{$^{\@thefnmark}$\hss}}
 \if@twocolumn
 \twocolumn[\@maketitle]
 \else \newpage
 \global\@topnum\z@ \@maketitle \fi\thispagestyle{firstpage}\@thanks
 \endgroup
 \setcounter{footnote}{0}
 \let\maketitle\relax
 \let\@maketitle\relax
 \gdef\@thanks{}\gdef\@author{}\gdef\@title{}\let\thanks\relax}
\makeatother

\def\O{{\Scr O}}
\def\C{{\mathbb C}}
\def\P{{\mathbb P}}

\def\R{{\mathbb R}}
\def\Z{{\mathbb Z}}

\def\Hom{\operatorname{Hom}}
\def\Ext{\operatorname{Ext}}

\def\ch{\operatorname{ch}}
\def\td{\operatorname{td}}

\def\SO{\operatorname{SO}}

\def\GO{\operatorname{O{}}}
\def\SU{\operatorname{SU}}
\def\GU{\operatorname{U{}}}

\def\rank{\operatorname{rank}}

\def\labto#1{\mathrel{\mathop\to^{#1}}}
\def\Dr{\mathbf{D}}

\def\sm{$\sigma$-model}
\def\nlsm{non-linear \sm}
\def\CY{Calabi--Yau}

\def\cM{{\Scr M}}

\def\cF{{\Scr F}}

\def\ff#1#2{{\textstyle\frac{#1}{#2}}}

\def\HS#1{{\mathbb{F}}_{#1}}

\begin{document}
\thispagestyle{empty}

\leftline{\copyright~ 1998 International Press}
\leftline{Adv. Theor. Math. Phys. {\bf 2} (1998) 1041-1074}  

\vspace{0.4in}
\begin{center}
{\huge \bf The Heterotic String,}

{\huge \bf The Tangent Bundle}

{\huge \bf  and Derived Categories}

\vspace{0.4in}

\({\bf  Paul\;   S.\; Aspinwall^{a}, \; Ron\; Y.\; Donagi^{b}}\)
\linebreak

$^{a}$Center for Geometry and Theoretical Physics,\\
Duke University, \\
Box 90318, \\
Durham, NC 27708-0318\\
$\:$  \\

$^{b}$ Department of Mathematics,\\
University of Pennsylvania,\\
 Philadelphia, PA 19104-6395\\




\vspace{0.2in}
\renewcommand{\thefootnote}{}
\footnotetext{\small e-print archive: {\texttt http://xxx.lanl.gov/abs/hep-th/9806094}}
\renewcommand{\thefootnote}{\arabic{footnote}}

{\bf Abstract} \\
\vspace{0.1in}
\parbox[c]{4.5 in}{\small {\hspace{0.2 in}}We consider the compactification of the $E_8\times E_8$ heterotic
string on a K3 surface with ``the spin connection embedded in the
gauge group'' and the dual picture in the type IIA string (or
F-theory) on a \CY\ threefold $X$.  It turns out that the same $X$
arises also as dual to a heterotic compactification on 24 point-like
instantons.  $X$ is necessarily singular, and we see that this
singularity allows the Ramond-Ramond moduli on $X$ to split into
distinct components, one containing the (dual of the heterotic)
tangent bundle, while another component contains the point-like
instantons.  As a practical application we derive the result that a
heterotic string compactified on the tangent bundle of a K3 with ADE
singularities acquires nonperturbatively enhanced gauge symmetry in
just the same fashion as a type IIA string on a singular K3 surface.
On a more philosophical level we discuss how it appears to be natural
to say that the heterotic string is compactified using an object in
the derived category of coherent sheaves.  This is necessary to
properly extend the notion of T-duality to the heterotic string on a
K3 surface.}

\end{center}

\vfil\break


\section{Introduction}    \label{s:int}

Back in the neolithic period of string theory \cite{CHSW:} it was
realized that one of the simplest methods of achieving a fairly realistic
model from string theory is to compactify the heterotic string on a
\CY\ manifold and ``embed the spin connection in the gauge
group''. Such a compactification consists of using the tangent bundle
of the \CY\ manifold as the vector bundle on which the gauge degrees
of freedom of the heterotic string are compactified. When this is
done, certain nonperturbative aspects of the world-sheet field theory
tend to simplify \cite{DSWW:}.
\pagenumbering{arabic}
\setcounter{page}{1042}

\pagestyle{myheadings}
\markboth{\it THE HETEROTIC STRING ...}{\it P. ASPINWALL, R. DONAGI}
Not long after this proposal, in such papers as \cite{Dist:res}, it
was realized that perhaps the tangent bundle did not play a
particularly distinguished
r\^ole in the compactification of heterotic strings. Having said that,
the underlying conformal field theory does have more supersymmetry in
the case of the tangent bundle than it would in the generic case and
for that reason alone one should expect it to have somewhat special properties.

More recently we have learnt that duality is a very powerful way of
probing all the nonperturbative effects that may come into play when
considering a compactification. In this context, the duality between
the heterotic string and the type IIA string (or F-theory) should be a
good way to analyze such compactifications.

At present the analysis of moduli for the heterotic string on a \CY\
threefold and of its duality with F-theory is a technically formidable
subject, although some progress has been made (see, for example,
\cite{FMW:F,BJPS:F,Don:spec,Don:F,BCGJL:F,CD:F4}). In this paper
we will tackle the
simpler question concerning the $E_8\times E_8$ heterotic string
compactified on the tangent bundle of a K3 surface. We embed the
$\SU(2)$ of the spin connection into one of the $E_8$'s leaving an
unbroken $E_7\times E_8$ generically. First, we wish to
pose the following question: On what \CY\ threefold $X$ should one
compactify F-theory so that the theory is dual to the $E_8\times
E_8$ heterotic string compactified on the tangent bundle of a K3
surface $S$? Equivalently one may find $X$ such that the type IIA
string compactified on $X$ is dual to the $E_8\times E_8$ heterotic string
compactified on the tangent bundle of the product of a K3 surface and
a 2-torus.

Although the study of the heterotic string on a K3 surface should be
much easier than for the heterotic string on a \CY\ threefold, it is
still far from completely understood. The initial conjectures
regarding its duality to the type IIA string were presented in
\cite{KV:N=2} and then progress was made in
\cite{MV:F,MV:F2,AG:sp32}.

If we go to part of the moduli space of theories where both the type
IIA string and the heterotic string are free from nonperturbative
corrections then one may define a systematic map between the two
theories by using ``stable degenerations'' as used in
\cite{FMW:F,AM:po,CD:F4,me:hyp}. This is the method by which we will attack
the tangent bundle problem.

The problem of finding the geometry of $X$ is not particularly difficult
but it gives rise to many questions. In particular we will find that
the tangent bundle appears to be almost identical to point-like
instantons. This raises a puzzle as the physics of the heterotic
string on the tangent bundle should be quite different than that of
point-like instantons.

We are led to a careful analysis of the Ramond-Ramond degrees of freedom
of the type IIA string, as it is these which distinguish the two
cases. This problem has also been studied for the case of the tangent
bundle in \cite{FMW:Ell}. We will discover that it is necessary
to make some refinements of the usual 
rules of F-theory in predicting gauge symmetries and counting of
tensor moduli in six dimensions.

The understanding of the tangent bundle can be pursued in various
directions. One of the things one may do is to build on the work of
papers such as \cite{AM:po,AM:frac} in cataloging the possibilities of
six dimensional physics for compactification of the heterotic string
on a K3 surface. We do this in section \ref{s:orb} and discover that
the heterotic string may acquire enhanced gauge symmetries on A-D-E
singularities in a way very similar to the type IIA string.

Instead, on a more fundamental level, we may try to analyze global questions
about the moduli space. As mentioned above, the usual
heterotic/F-theory duality arguments are confined to a large
codimension boundary of the moduli space where both theories are
``weakly-coupled''. We do not currently have a description of string
theory which is in any way exact as we move into the open
interior of the hypermultiplet moduli space --- both theories suffer
from quantum
corrections \cite{me:hyp}. By restricting to the tangent bundle of the
heterotic
string we will actually be able to penetrate deeply into the moduli
space along a closed subspace while retaining exactness. This will
lead us to some new claims about the definition of the data specifying
the heterotic string. We discuss this in section \ref{s:T}.

These two goals will require a fairly sophisticated
analysis of the geometry of sheaves on a K3 surface and this comprises
a large part of this paper. In particular in section \ref{s:FM} we
will review the mathematics of sheaves which we will require. Of
particular interest are the Mukai vector and the notion of the derived
category of coherent sheaves. In section \ref{s:hetT} we discuss how
the heterotic string compactified on a sheaf is interpreted in the type
IIA or F-theory language. We discuss how F-theory rules for
determining the massless spectrum are affected by RR moduli. We also
propose that a point-like instanton should be identified with the
ideal sheaf of a point. To some extent the contents of section \ref{s:FM} are
important for section \ref{s:T} and the contents of section
\ref{s:hetT} are important for section \ref{s:orb}. Having said this,
these two subjects are somewhat interconnected and it would be awkward
to disentangle them.


\section{The Fourier-Mukai Functor}  \label{s:FM}

\subsection{Some Sheaf Generalities}   \label{s:sh}

\def\pbun{\Scr{P}}
\def\shE{\Scr{E}}
\def\shF{\Scr{F}}
\def\Mv{\boldsymbol{\mu}}

To do the analysis required in this paper it will be much easier to use
the language of sheaves	rather than vector bundles.
This will allow us to handle vector bundles on the K3 surface, vector
bundles over curves within this surface, and ``skyscraper'' sheaves all at
once.
Consider a vector
bundle $V$ over some manifold $M$.
We will assume that the vector bundle is a holomorphic vector bundle,
and in particular that the structure group of the bundle is $\SU(N)$ for
some $N$. This assumption is valid for what we require in this paper
but is certainly not sufficient for a general analysis of the
heterotic string. See \cite{Don:spec,Don:F,CD:F4,me:hyp} for methods
in the general case.
The sheaf version of this holomorphic vector bundle is
to associate to each open neighbourhood $U\subset M$, the group of
holomorphic sections of $V$ over $U$. A sheaf obtained in this way
is called ``locally free''. There are sheaves which are not locally
free however. Consider a submanifold $B\subset M$. One may consider a
sheaf over $M$ which is defined by a vector bundle $W\to B$ as
follows. For every open neighbourhood $U\subset M$ associate the group
of holomorphic sections of $W$ restricted to $U\cap B$. If $U\cap B$
is empty, take this group to be 0.

Thus we can have a sheaf over $M$ defined by a vector bundle over some
subspace of $M$. In a sense, the rank of this vector bundle has some
fixed value over $B$ but is zero elsewhere. Indeed in this way a sheaf
can represent a bundle whose rank varies over $M$. In the example
where the groups associated to the sheaf vanish unless $U\cap B$ is
nonempty, then the sheaf is said to be ``supported'' over $B$. The
extreme example of this is the ``skyscraper sheaf'' which is supported
only at a single point, where the ``fibre'' is $\C$.

Sheaves are very natural objects in the context of string
duality. They have been used for example in \cite{DGM:sh,HM:alg}. It
seems clear that one cannot hope to gain a full understanding of the
moduli space of string theories when restricting one's attention to only
locally-free sheaves. We will also see further evidence in this paper
to this effect. The main difference between vector bundles and sheaves is
that the former objects can be described in terms of differential
geometry, the latter objects are creatures very much of an
algebro-geometrical upbringing. For this reason, much of the
mathematics involved tends to be of an algebraic nature.

If we have some algebraic variety $X$, the most basic sheaf over $X$
is the structure sheaf $\O_X$. This is the sheaf built from sections
of the trivial holomorphic line bundle over $X$ if $X$ is smooth. One
can view $\O_X$ as a ring. It is then natural to generalize the
concept of a holomorphic vector bundle to an $\O_X$-module. A free
$\O_X$-module will correspond to a trivial vector bundle. We will assume
that we are always dealing with a particular type of $\O_X$-module ---
namely a {\em coherent\/} sheaf.\footnote{If one believes that string
theory has an algebraic origin then (quasi-)coherent sheaves are a natural
choice. A (quasi-)coherent sheaf is defined as a sheaf which may be
constructed from the same underlying algebraic structure as the scheme
on which it lives. See \cite{Hartshorne:} for an exact statement of
this.}

We will be concerned with the way that one may build sheaves from
others using the algebraic notion of ``extensions''. Let $\shE_1$ and
$\shE_2$ be two sheaves over some fixed variety $X$. To be pedantic,
we consider $\shE_1$ and $\shE_2$ to be $\O_X$-modules. One may then
build a third sheaf, $\shF$, from the exact sequence of
$\O_X$-modules:
\begin{equation}
  0 \to \shE_1 \labto a \shF \to \shE_2 \to 0.   \label{eq:es1}
\end{equation}
Here, $\shF$ is not defined uniquely. Algebraically, in the above
sequence $\shF$ is defined by an element of
$\Ext_{\O_X}(\shE_2,\shE_1)$.
As a sheaf however, two elements of
$\Ext_{\O_X}(\shE_2,\shE_1)$ which differ by a nonzero constant give
rise to two isomorphic
$\shF$'s. This is because we may multiply the map $a$ in
(\ref{eq:es1}) by a nonzero constant, thus changing the extension
class without changing the sheaf $\shF$.

One must pay special attention to rather peculiar things which may
happen when considering the moduli space of sheaves (or indeed vector
bundles). Of particular
concern is the notion of ``$S$-equivalence''. Let us consider the
problem of building a family of sheaves of the above form over some
fixed variety $X$. Let $A=\Ext_{\O_X}(\shE_2,\shE_1)$. We may then
build the ``universal extension'', $\shF_A$ over $A\times X$ by
\begin{equation}
  0 \to p_2^*\shE_1 \to \shF_A \to p_2^*\shE_2 \to 0,
\end{equation}
where $p_2:A\times X\to X$ is the natural projection. The restriction
of $\shF_A$ to a point $a\in A$ will then specify a sheaf $\shF_a$
over $X$.

Consider a complex line through the origin in $A$ and restrict
$\shF_A$ to it. From what we have said above $\shF_a$ will be
everywhere the same except where $a=0$. That is we have a family of
sheaves which is constant everywhere except for one member where it
``jumps'' to something inequivalent. This is precisely the kind of
thing one wants to avoid when building a nice moduli space!

The simplest approach is to define our way out of the above problem by
saying that we wish to build the moduli space of $S$-equivalence
classes of sheaves over $X$. We then say that two sheaves $\shF_1$ and
$\shF_2$ are ``$S$-equivalent'' if we may build a family of sheaves
which are all isomorphic to $\shF_1$ except for one member which is
isomorphic to $\shF_2$.

While such an approach is fine 
if one is concerned with the isolated problem of building a nice-looking
moduli space, we are supposed to be describing moduli spaces of string
theories in this paper. Presumably string theory ``knows'' whether it
really wants $S$-equivalent but not isomorphic sheaves to be
considered the same or not.

An example which will be important to us is the following. Let $X=E$,
an elliptic curve, and let $\shE_1$ and $\shE_2$ be isomorphic to
$\O_E$. Thus we have
\begin{equation}
  0 \to \O_E \to \shF \to \O_E \to 0,  \label{eq:OFO}
\end{equation}
where $\shF$ is a rank 2 sheaf. Now $\Ext_{\O_E}(\O_E,\O_E) \cong
H^1(E,\O_E)\cong\C$ (see for example page 234 of \cite{Hartshorne:})
and so (\ref{eq:OFO}) defines two possibilities for $\shF$. Either we
have the trivial extension $\cF\cong\O_E\oplus\O_E$, or we have the
unique nontrivial extension.

Consider now building the moduli space of degree zero rank 2
(semi-stable) sheaves over $E$. The moduli space of $S$-equivalence
classes will be nice but the true moduli space of sheaves will
essentially have two points, representing the two possibilities above,
at one of the locations of moduli space --- the moduli space is not
separated. This is discussed in some detail in \cite{Don:F}.

Now if we were to consider the moduli space of heterotic strings on a
2-torus, the above system should fit nicely into it. The moduli space
of the heterotic string on a 2-torus is well-understood and has no
pathologies. Thus in this case, we need to use the moduli space of
$S$-equivalence classes to get agreement with string theory.
This can be explained because in this case the moduli space
in question can be written in terms of the moduli space of solutions to
Yang--Mill's equations. The relationship between Yang--Mill's bundles
and stability is well-understood (see for example \cite{AtBot:YM}). As we
shall see however, it is not true that we may always use
$S$-equivalence for all problems in string theory. The distinction
becomes especially important when we consider families of sheaves:
the behavior of the family away from the central fiber may force us
to choose one of several $S$-equivalent objects over the special fiber.
In section
\ref{ss:tsc} we will in fact want to distinguish between the two
inequivalent extensions of (\ref{eq:OFO}).


\subsection{The Fourier-Mukai Transform of a Sheaf}  \label{ss:FM}

In this section we wish to analyze the moduli space of sheaves
on an elliptic K3 surface. We will review the usual method
\cite{FMW:F,Don:F,BJPS:F}
of first analyzing sheaves on a single elliptic curve and then
extending this idea to elliptic surfaces.

Let $E$ be a 2-torus with a complex structure and a distinguished
point $0\in E$. That is, $E$ is an elliptic curve. The moduli space of
flat line bundles on $E$ may be viewed as the space of flat
connections which in turn can be viewed as the space
$\hat E=H^1(E,\R/\Z)$. $\hat E$ is dual to $E$. We may also view $\hat
E$ as the Jacobian of $E$. Actually
$\hat E$ has the same complex structure as $E$ and so is in some sense
isomorphic to $E$. Indeed we may define a degree zero line bundle on
$E$ by fixing a meromorphic section which has a pole at $0$ and a zero
at some other point $x\in E$. The moduli space of line bundles on $E$
of degree zero
may then be identified with the moduli space of $x$ --- which is $E$.

The Poincar\'e bundle, $\pbun$, is a bundle over $E\times\hat E$ with
the following properties. It is ``universal'' in the sense that when
restricted to $E\times x$, where $x\in\hat E$, it gives the bundle over
$E$ defined by the point in the moduli space $x\in\hat E$. We also
demand that $\pbun$ restricted to $0\times\hat E$ is trivial.
We mainly view $\pbun$ as the associated sheaf to this vector
bundle. Formally we may define $\pbun$ as
\begin{equation}
  \pbun = \O_{E\times\hat E}(\Delta-0\times\hat E-E\times\hat0),
\end{equation}
where $\Delta$ is the diagonal divisor representing the graph of the
isomorphism $E\to\hat E$.

Let $Z$ be a surface and let $\pi:Z\to B$ be an elliptic fibration. We
may replace every fibre by its Jacobian to obtain
another elliptically fibred surface $\hat\pi:\hat Z\to B$. Some care
is needed in analysis of this process over the bad fibres (see, for
example, \cite{CDol:EnI}). If the fibration $\pi:Z\to B$ has a
section, which we assume is the case from now on, then $\hat Z$ is
actually isomorphic, as a complex variety, to $Z$. Nevertheless we
will often find it convenient to distinguish between $Z$ and $\hat Z$,
as was done in \cite{BJPS:F}.

We may extend $\pbun$ to be a sheaf over $Z\times_B\hat Z$ in the
obvious way (except again for subtleties at bad fibres). Recall that
$Z\times_B\hat Z$ is the fibration
$Z\times_B\hat Z\to B$ with fibre $\pi^{-1}(b)\times\hat\pi^{-1}(b)$
for any point $b\in B$. We may also define projection maps
$p:Z\times_B\hat Z\to Z$ and $\hat p:Z\times_B\hat Z\to \hat Z$.

Now take a nice smooth locally-free sheaf, $\cF$, associated to some
smooth $\SU(N)$-bundle over $Z$. Consider the following sheaf:
\begin{equation}
  F^1(\shF) = R^1\hat p_*(\pbun\otimes p^*\shF).  \label{eq:FM0}
\end{equation}
It is common to call $F^1(\shF)$, which is a sheaf over $\hat Z$, the
``Fourier-Mukai transform'' of $\shF$ after \cite{Muk:FM}.
We are working in the ``relative'' setting. That is we are applying
the transformation to each elliptic fibre of a fibration. This was
analyzed in \cite{BBRP:FM}. We refer to this paper for more details of
what follows. We now explain (\ref{eq:FM0}) in detail.

For any sheaf $\shE$ over $Z\times_B\hat Z$, one may loosely take
$R^1\hat p_*\shE$ to be the sheaf whose ``fibre'' (or stalk) over a point
$z\in\hat Z$ is the group $H^1(E_z,\shE|_{E_z})$, where $E_z\subset Z$
is the elliptic fibre $\pi^{-1}\hat\pi(z)$.  Let $\shE = \pbun\otimes
p^*\shF$.  Since we have an $\SU(N)$-bundle (with zero $c_1$) the
sheaf $\shE|_{E_z}$ will have degree zero. Riemann-Roch then tells us
that $\dim H^0(E_z,\shE|_{E_z}) = \dim H^1(E_z,\shE|_{E_z})$. Now a
generic degree zero sheaf over $E_z$ will have no global sections and
so $H^1(E_z,\shE|_{E_z})=0$ generically. This shows that $F^1(\shF)$
is supported over some proper subset of $\hat Z$.

The sheaf $F^1(\shF)$ will be supported at points where $\shE|_{E_z}$
contains a trivial summand. This will generically happen over $N$
points of the elliptic curve $\hat\pi^{-1}(b)$ where the Poincar\'e
bundle nicely cancels out one of the summands of $\shF$. Therefore
$F^1(\shF)$ is supported on a curve which is an $N$-fold cover of
$B$. This is the ``spectral curve'' $C_S\subset\hat Z$. Generically only
one summand will be trivialized at a time and so $F^1(\shF)$ is rank
one over $C_S$.

As an example let us consider the simplest case of $\O_Z$. That is, we
begin with the sheaf corresponding to the trivial line bundle over
$Z$. Now clearly $F^1(\O_Z)=R^1\hat p_*\pbun$. This will be a sheaf
supported along the section of the fibration $\pi:Z\to B$. Actually, it
follows from either relative duality or from Grothendieck-Riemann-Roch
(see, for example Theorem 2.8 of \cite{BBRP:FM}) that this
sheaf corresponds to the canonical bundle of $B$, i.e., a degree $-2$
line bundle over $B\cong\P^1$.

It is natural to consider a closely-related transform:
\begin{equation}
  F^0(\shF) = \hat p_*(\pbun\otimes p^*\shF).
\end{equation}
In our above argument this would amount to considering a sheaf whose
stalks were given by $H^0(E_z,\shE|_{E_z})$ rather than
$H^1(E_z,\shE|_{E_z})$. One might believe therefore that it is equivalent
to $F^1(\shF)$ in some sense. Actually, due to sloppiness in the above
argument this is not the case. One really should be very careful when
building these sheaves to treat the notion of a stalk of a sheaf
correctly. If $\shF$ is a locally-free sheaf as above then
$F^0(\shF)=0$. This follows from the fact that sections of $F^0(\shF$) over
an open set $U\subset\hat Z$ are the same as sections of $\pbun\otimes
p^*\shF$ over $\hat p^{-1}(U)$. Since fibres of the latter vanish at
most points of $\hat p^{-1}(U)$, the sections must all vanish identically.

The reasons for the difference between $F^0$ and $F^1$ are important
but rather complicated and we do not discuss them here. We refer the
reader to section III.12 of \cite{Hartshorne:} for a full treatment.

Now let $\shE_0$ be a sheaf on $Z$ supported over the zero section
where it is the trivial line bundle, of degree zero. It is not hard to
see that $F^0(\shE_0)\cong\O_Z$. Thus, except for the mismatch in
degrees, $F^0$ looks like the inverse of $F^1$ when we act on the
structure sheaf $\O_Z$.

To cure this we introduce a new transform:
\begin{equation}
  T(\shF) = \shF\otimes\O_Z(E),
\end{equation}
where $\O_Z(E)$ is the sheaf corresponding to the line bundle whose
$c_1$ is equivalent to a generic elliptic fibre of $Z$. We also 
define:\footnote{
Note that our use
of $S$ and $T$ differs from \cite{BBRP:FM}. Our notation is supposed to
be reminiscent of $SL(2,\Z)$ in that $S$ satisfies $S^2=-1$ while $T$
is of infinite order. But actually our $S$ and $T$ sit in the two distinct
copies of $SL(2,\Z)$ (inside $O(2,2;\Z)$), so they commute, and generate
an abelian subgroup $\Z \times \Z_4$.)
}
\begin{equation}
  S^i(\shF) = TF^i(\shF)
    = R^i\hat p_*(\pbun\otimes p^*\shF)\otimes\O_{\hat Z}(E).
\end{equation}

The reader may like to check that for a locally-free sheaf $\shF$
with $c_1=0$, $S^0S^1(\shF)$ is very
nearly the same thing as $\shF$. That is $S^0$ acts rather like the
inverse of $S^1$. The slight mismatch is that $S^0S^1$ actually
acts as $-1$ on every elliptic fibre of $\pi:Z\to B$.
Since $S^0$ takes a locally free sheaf to 0, then it cannot be
generally true that $S^0$ and $S^1$ are inverses to each other for any
sheaf. In order to make
a cleaner statement about $S^0$ and $S^1$ we are required to go to the
``derived category''.


\subsection{The Derived Category and the Mukai Vector} \label{ss:der}

Our goal is to find some kind of transformation we can do to a sheaf
to turn it into another sheaf such that the resulting sheaf is
``equivalent'' to the first sheaf as far as string theory is
concerned.

The $S^1$ generator above looks promising in as much as it gives us a
sheaf supported along the spectral curve when applied to a locally
free sheaf. It is somewhat unsatisfactory that $S^1$, when applied to
this resulting sheaf on the spectral curve, gives us 0. Clearly $S^1$
does not generally map within any kind of nontrivial equivalence class.
We had
to use $S^0$ instead to return to our original locally-free sheaf. It
would be useful if we could somehow combine $S^0$ and $S^1$ into a
single operator $\mathbf{S}$. (We also combine $F^0$ and $F^1$ into
$\mathbf{F}$.)

In order to do this we need to change the objects we use.
It turns out that rather than thinking about sheaves, we should think of
complexes of sheaves. In categorical language we will use the ``derived''
category of coherent sheaves. We will denote the (bounded) derived
category of coherent sheaves on $Z$ by $\Dr(Z)$.
The notion of derived categories is
rather complicated and the details are beyond what we require for this
paper. We refer the reader to \cite{Hart:dC} for an account of the
original motivation of their construction or \cite{Wei:hom} for a
detailed definition.

The general idea is this. An object of the derived category is no longer
a sheaf $\shF$, but rather a complex of sheaves $\mathbf{C}(\shF)$:
\begin{equation}
  \ldots \to C^2(\shF) \to C^1(\shF) \to C^0(\shF) \to \ldots,
\end{equation}
taken up to certain equivalences.
For example, if one begins with a sheaf $\shF$, then one
may associate to this the complex $\mathbf{S}(\shF)$:
\begin{equation}
\ldots \to 0 \to S^1(\shF) \stackrel{0}{\to} S^0(\shF) \to 0 \to \ldots
\end{equation}
Often only one of the terms in this complex will be nontrivial and we
may use this sheaf to ``represent'' $\mathbf{S}(\shF)$. When this is
the case, the functor $\mathbf{S}$ looks like it should satisfy our
requirements. In general however $\mathbf{S}$ is a functor acting on
the {\em derived\/} category of sheaves and not the category of
sheaves itself.

If we identify $\hat Z$ with $Z$ we may then state the following
relative version of Mukai's result.
\begin{theorem}
The functor $\mathbf{S}$ satisfies
\begin{equation}
  \mathbf{S}^2 = (-1)_E[-1],   \label{eq:T2}
\end{equation}
where $(-1)_E$ is the inversion of each elliptic fibre and $[-1]$
denotes a shift of the complex of one to the right. \label{th:1}
\end{theorem}
We refer to \cite{Muk:FM,BBRP:FM} for a proof.

The other construction due to Mukai that we will also find of great
use in this paper is the ``Mukai Vector'' \cite{Muk:bun}. Let $Z$ be a
K3 surface and let us represent an element of
$H^*(Z,\Z)=H^0(Z,\Z)\oplus H^2(Z,\Z)\oplus H^4(Z,\Z)$ by the triple
$(a,b,c)$. Here $a\in H^0(Z,\Z)$, $b\in H^2(Z,\Z)$, and $c\in
H^4(Z,\Z)$. We may take $a$ and $c$ to be integers since $H^0(Z,\Z)
\cong H^4(Z,\Z)\cong\Z$.

If $\shF$ is a coherent sheaf we then define the Mukai vector
$\Mv(\shF)\in H^*(Z,\Z)$ by
\begin{equation}
\begin{split}
  \Mv(\shF) &= \ch(\shF).\sqrt{\td(Z)} \\
      &= \Bigl(r,c_1(\shF),r +\ff12(c_1^2(\shF)-2c_2(\shF))\Bigr),
\end{split}
\end{equation}
where $r$ is the rank of $\shF$.
The Mukai vector has already made appearances in the string
literature. It has appeared in the context of D-branes and anomalies
in \cite{GHM:inf} and has been used subsequently in work such as
\cite{HM:alg}. Some interesting observations about the nature of the
Mukai vector in string theory were also made in \cite{MM:K}.
Given that it takes values in $H^*(Z,\Z)$ which is a
very natural object in string theory \cite{AM:K3p}, it is not at all
surprising that it should also make an appearance in the analysis of
the heterotic string. Note that $\Mv$ has the following property
induced from the Chern character:
\begin{equation}
  \Mv(\shE\oplus\shF) = \Mv(\shE) + \Mv(\shF).
\end{equation}

We may also extend the definition of $\Mv$ to $\Dr(Z)$ by
\begin{equation}
  \ch(\mathbf{C}(\shF)) = \sum_i(-1)^i\ch(C^i(\shF)).   \label{eq:chdC}
\end{equation}
By doing this we can specify the induced action of $\mathbf{S}$ on $H^*(Z,\Z)$.
By using (\ref{eq:T2}) we can immediately see the induced action of
$\mathbf{S}^2$. Since $(-1)_E$ acts trivially on our even homology
cycles and $[-1]$ reverses the sign because of (\ref{eq:chdC}), we see
that $\mathbf{S}^2$ simply acts as $-1$ on $H^*(Z,\Z)$.

We will consider the case where $Z$ is a generic elliptic K3 surface
with a section. This means that the Picard lattice of $Z$ is a two
dimensional lattice in the form of a hyperbolic plane $U$. Let $v_0$
represent the class of the generic elliptic fibre and let $v_1$
represent the sum of the class of the section and $v_0$. It can be easily
seen that the intersection form on $Z$ gives rise to the product
\begin{equation}
\begin{split}
  v_0.v_0 &= v_1.v_1 = 0\\
  v_0.v_1 &= 1.
\end{split}
\end{equation}
We also define an inner product on $H^*(Z,\Z)$ by
\begin{equation}
  (a_1,b_1,c_1).(a_2,b_2,c_2) = b_1.b_2 - a_1c_2 - c_1a_2.
\end{equation}

Before we proceed we need one more fact. Consider a line bundle of
degree d on some curve $C\subset Z$. Let $\shF$ be the sheaf on $Z$
but supported only on $C$ corresponding to this bundle. One may show that
\begin{equation}
\begin{split}
  c_1(\shF) &= C \\
  c_2(\shF) &= C.C - d,
\end{split}
\end{equation}
where $C.C$ is the self-intersection of $C$.

We now specify the action of $\mathbf{S}$ on some elements of
$H^*(Z,\Z)$.
\begin{itemize}
  \item $\Mv=(1,0,1)$. This is the structure sheaf $\O_Z$.
As mentioned above, $F^1$ transforms this to a sheaf of degree $-2$
supported along the section $v_1-v_0$. Applying $T$ will change the
degree to $-1$. It follows that
\begin{equation}
  \mathbf{S}(1,0,1) = (0,v_0-v_1,0).
\end{equation}
  \item $\Mv=(0,v_0,0)$. This is a sheaf supported along a single
elliptic fibre of $Z$. Over this fibre, the sheaf has rank one and
degree 0. When $S^1$ is applied, we obtain the skyscraper sheaf
$\O_z$ supported at the single point $z$. Applying $T$ does
nothing. One may show that $c_2(\O_z)=-1$ and therefore
\begin{equation}
  \mathbf{S}(0,v_0,0) = (0,0,-1).
\end{equation}
  \item $\Mv=(0,v_1-v_0,0)$. The corresponding sheaf with this Mukai
vector is $\O(-1)$ on the section. Its $F^0$ is $\O_Z(-E)$, so its
$S^0$ is $\O_Z$, and we have:
\begin{equation}
  \mathbf{S}(0,v_1-v_0,0) = (1,0,1).
\end{equation}
  \item $\Mv=(0,0,1)$. This is the skyscraper sheaf $\O_z$. Applying
$S^0$ to this we obtain the sheaf supported along a single elliptic
fibre. Again $T$ does nothing.
\begin{equation}
  \mathbf{S}(0,0,1) = (0,v_0,0).
\end{equation}
\end{itemize}

To review (and using linearity) we obtain
\begin{equation}
\begin{split}
  \mathbf{S}(1,0,0) &= (0,-v_1,0) \\
  \mathbf{S}(0,v_0,0) &= (0,0,-1) \\
  \mathbf{S}(0,v_1,0) &= (1,0,0) \\
  \mathbf{S}(0,0,1) &= (0,v_0,0)
\end{split}
\end{equation}

Note that $\mathbf{S}^2=-1$ as required by theorem \ref{th:1} and
that, up to signs,
$\mathbf{S}$ simply exchanges the hyperbolic plane generated by
$(v_0,v_1)$ with the hyperbolic plane generated by $H^0(Z)$ and
$H^4(Z)$.

As an application let us consider a locally-free sheaf, $\shF_k$, corresponding
to an $SU(N)$-bundle with $c_2=k$ over a K3 surface $Z$. The above
shows that
\begin{equation}
\begin{split}
  \Mv(\shF_k) &= (N,0,N-k)\\
  \Mv(\mathbf{F}(\shF_k)) &= (0,(N-k)v_0-Nv_1,N)\\
  \Mv(\mathbf{S}(\shF_k)) &= (0,(N-k)v_0-Nv_1,0).
\end{split}
\end{equation}
This is consistent with the fact that $F^1(\shF_k)$ is a sheaf
supported on a curve in the class $(k-N)v_0+Nv_1$ --- the spectral
curve $C_S$. This curve has self-intersection $2N(k-N)$ and thus has
genus $g=N(k-N)+1$. It follows from the above calculations that
the degree of the line bundle on $C_S$ producing the required sheaf is
$N(k-N-1)=g-1-N$.\footnote{There appears to be a typo in \cite{BJPS:F}
for this formula.}
This result can also be obtained by noting that the direct
image $\hat{\pi}_*$ preserves Euler characteristics: in our case, the
image $\hat{\pi}_*(F^1(\shF_k))$ is a rank
$N$ vector bundle on $B=\P^1$, which is easily seen to be the canonical
bundle $\O_{\P^1}(-2)$ tensored with the restriction of $\shF_k$.
Its Euler characteristic therefore equals $-N$. This agrees with
the Euler characteristic of a line bundle of degree $g-1-N$ on $C_S$.

We end this section by noting that one of the uses of the Mukai vector
is to calculate the (complex) dimension of
the moduli space of sheaves. To do this we restrict to the case where
$\shF$ is ``simple'' --- that is the only automorphisms of $\shF$ are
global rescalings. In this case the dimension of the moduli space is
given by \cite{Muk:bun}
\begin{equation}
  \dim\Ext_{\O_Z}(\shF,\shF) = \Mv(\shF). \Mv(\shF) +2.
\end{equation}

This restriction to simple sheaves is fairly severe but will work
for the above locally free sheaf $\shF_k$. In this case the dimension of the
moduli space is $2g=2N(k-N)+2$ in agreement with standard methods.


\section{The heterotic string on the tangent bundle}  \label{s:hetT}

\subsection{Review}  \label{ss:rev}

The general system we wish to analyze is an $E_8\times E_8$ heterotic
string compactified on a product of a K3 surface, $Z$, and a
2-torus. This is supposedly dual to the type IIA string compactified
on a \CY\ threefold, $X$. We refer the reader to \cite{me:lK3,Don:HetF,me:hyp}
for a general description of this system together with much of the
notation we will be using.

We will study the moduli space coming from the heterotic string
corresponding to deformations of $Z$ together with a sheaf on
$Z$. This maps by duality to deformations of the complex structure of
$X$ together
with ``RR''-parameters which, when $X$ is smooth, take values in
$H^3(X,\R/\Z)$ (and the
dilaton-axion). This part of the moduli space has a quaternionic
K\"ahler structure and is the moduli space of hypermultiplets. The
heterotic string's 2-torus has nothing to do with this moduli space
and can be ignored. Thus we may view the physics as derived from a
six-dimensional theory obtained by compactifying the heterotic string
on $Z$. This is the F-theory picture.

Let us fix notation. We will assume that $X$ is a \CY\ threefold which
has the structure both of an elliptic fibration $\pi_F:X\to\Sigma$ and
a K3 fibration $p_F:X\to B$. Here $\Sigma$ is a Hirzebruch surface
$\HS n$ and $B\cong\P^1$. $Z$ is an elliptic fibration $\pi:Z\to B$ as
above. All the fibrations have sections. See \cite{me:lK3} for a
discussion of these assumptions.

Let $C_0$ be the section of $\Sigma$ with self-intersection
$-n$. Thus, $C_0$ is the isolated section of $\Sigma$ if $n>0$. Let
$f$ be a generic $\P^1$-fibre of $\Sigma$. Let $\sigma_0$ be the
section of $\pi:Z\to B$ and let $E$ refer to a generic elliptic fibre
of either $\pi:Z\to B$ or $\pi_F:X\to\Sigma$. Connecting to section
\ref{ss:der} we see that $v_0=[E]$ and $v_1=[\sigma_0]+[E]$.

One of the key points is that neither the heterotic nor type IIA
string description of the moduli space is exact. Both suffer from
quantum corrections. In the type IIA string case, we may switch the
corrections off by setting the dilaton equal to $-\infty$. In terms of
the heterotic string we may take this limit by letting the area of
$\sigma_0$ go to infinity.

Once this is done we may remove the quantum corrections
from the heterotic side by making the generic fibre of this elliptic
fibration very large.
As we will see later, this is not a necessary condition to remove the
corrections but it is sufficient.
In terms of $X$, this deformation corresponds to a change in
complex structure which takes $X$ to a stable degeneration. This
degeneration is of the form of two threefolds, $X_1$ and $X_2$ which
intersect along a K3 surface isomorphic to $Z$.
We identify $Z$ with this intersection from now on.

Essentially string duality identifies the geometry of $X_1$ with one
of the $E_8$'s of the heterotic string and $X_2$ with the other
$E_8$. From now on we just focus on $X_1$ --- all things stated also
being true for $X_2$. We now have a fibration $p^1_F:X_1\to B$ with
fibre a rational elliptic surface and an elliptic fibration
$\pi^1_F:X_1\to\Sigma$ where $\Sigma$ is still the Hirzebruch surface
$\HS n$ \cite{AM:po}. We let $C_*$ denote the section of $\Sigma$
disjoint from $C_0$ where $X_1$ intersects $X_2$. Thus $C_*$ is really
the same thing as $\sigma_0$. Note that $C_*$ is isolated if $n<0$ (as
will happen the case of interest).

The elliptic fibration $\pi^1_F:X_1\to\Sigma$ degenerates over a
discriminant curve $\Delta^1\subset\Sigma$. The geometry of $X_1$ also
determines a spectral curve $C_S^1\subset Z$. Indeed the moduli of
$\Delta^1$ determine the moduli of $C_S^1$. We refer to \cite{me:hyp}
for a description of this.

Under the stable degeneration the deformations of complex structure
and RR-parameters on $X$ divide into three pieces \cite{me:hyp}. The
complex structure of $X$ defines the complex structure of $Z$ and the
complex structures of $C_S^1$ and $C_S^2$, together with the embeddings,
$C_S^1\subset Z$, $C_S^2\subset Z$. The RR-parameters of $X$ describe
a $B$-field in $H^2(Z,\R/\Z)$ and a line bundle (of a fixed degree) on
each of the spectral curves by specifying an element of the Jacobian
$H^1(C_S,\R/\Z)$.

Suppose now we have a bundle with structure group $SU(N)\subset E_8$
used to compactify one of the $E_8$'s of the heterotic string. In the
language of the last section, the data for the spectral curve, $C_S$, and its
line bundle is simply obtained by applying $F^1$ (or $S^1$) to the sheaf
corresponding to this vector bundle.

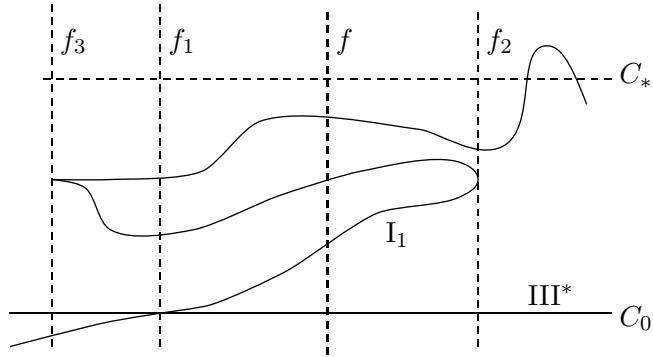
\begin{figure}
\begin{center}
\setlength{\unitlength}{0.008750in}%
\begin{picture}(365,210)(140,575)
\thinlines
\put(140,600){\line( 1, 0){360}}
\multiput(330,780)(0.00000,-8.03922){26}{\line( 0,-1){  4.020}}
\multiput(230,580)(0.00000,8.03922){26}{\line( 0, 1){  4.020}}
\multiput(420,780)(0.00000,-7.84314){26}{\line( 0,-1){  3.922}}
\multiput(160,740)(8.00000,0.00000){43}{\line( 1, 0){  4.000}}
\multiput(165,580)(0.00000,8.03922){26}{\line( 0, 1){  4.020}}
\put(140,580){
}
\put(165,680){
}
\put(425,755){\makebox(0,0)[lb]{$f_2$}}
\put(505,590){\makebox(0,0)[lb]{$C_0$}}
\put(505,735){\makebox(0,0)[lb]{$C_*$}}
\put(450,605){\makebox(0,0)[lb]{III$^*$}}
\put(235,755){\makebox(0,0)[lb]{$f_1$}}
\put(335,755){\makebox(0,0)[lb]{$f$}}
\put(170,755){\makebox(0,0)[lb]{$f_3$}}
\put(365,640){\makebox(0,0)[lb]{I$_1$}}
\end{picture}
\end{center}
  \caption{The discriminant for an SU(2)-bundle.}
\label{fig:SU2}
\end{figure}

Of particular interest to us is the case of $\SU(2)$-bundles with
$c_2=k$ . The geometry of $\Sigma$ is then fixed by $n=12-k$
\cite{MV:F}. The
geometry of this case was explicitly constructed in section 3.2 of
\cite{me:hyp}.
We show the geometry of $\Delta^1\subset\Sigma$ in figure
\ref{fig:SU2}. $\Delta^1$ is a reducible curve having a component
along $C_0$ above which the elliptic fibres are in Kodaira class
III$^*$ and another component above which the class is I$_1$.
Here $S_C$ is a double cover of $B$ branched at $4k-4$
points. Mapping $\Sigma$ to $B$, these branch points occur at $k-4$
places where the I$_1$
component of $\Delta^1$ collides with $C_0$ (marked in figure
\ref{fig:SU2} as $f_1$) and the $3k$ points of tangency of $\Delta^1$
with the $f$ direction (marked as $f_2$).

Generically, $X_1$ has a curve of $E_7$ singularities (from the
III$^*$ fibres) which
suggests an $E_7$ gauge symmetry \cite{MV:F2}. This arises since $\SU(2)\subset
E_8$ centralizes $E_7$. In some special cases this need not be the
exact gauge symmetry as we discuss later.


\subsection{Two special cases}  \label{ss:tsc}

\def\PIsh{\Scr{Z}_{24}}

{}From above we see that for a nice smooth $\SU(2)$-bundle, or the
associated sheaf $\shF_k$, we obtain a
spectral curve $C_S^1\subset Z$ in the class
\begin{equation}
  [C_S^1] = 2[\sigma_0]+k[E].   \label{eq:Ccl}
\end{equation}
Over this curve we have a line bundle of degree $2k-6$. This represents
the sheaf $F^1\shF_k$. The moduli of the curve $C_S^1$ are fixed, in the
type IIA language, by the moduli of $X_1$. The moduli of the line
bundle are fixed by the RR-fields in the type IIA string.

If $C_S^1$ is smooth then the moduli space of line bundles is a torus
(abelian variety) $H^1(C_S^1,\R/\Z)$ parametrized in the type IIA
language by elements of $H^3(X_1,\R/\Z)$. We will be concerned with the
case where $C_S^1$ and $X_1$ are {\em not\/} smooth. In this case, the
Jacobian of $C_S^1$ need not be a torus and so the RR-fields will not
parameterize a torus either.

In the genus one case, this degeneration of the Jacobian is familiar
from Kodaira's classification. We are doing a similar construction
here except that the spectral curve has a large genus. Thus, we want
to know how an abelian variety of a large complex dimension can
degenerate in a family. This is in general a very hard question. In
our case we will see that the Jacobian becomes reducible, so the RR
moduli can take their values in either of several distinct
components. Quite generally, this happens whenever the spectral curve
becomes non-reduced. A detailed analysis of these components is given
in \cite{DEL:H} in the case where the spectral cover becomes
everywhere non-reduced $C=2C'$. In our present situation there is one
additional feature, namely that only part of the spectral curve
becomes non-reduced: it becomes (twice) the zero-section $\sigma_0$,
plus (once) each of 24 fibres. However, the distinction we will
establish between the two F-theory compactifications will be entirely
due to the contribution of the non-reduced locus as seen in \cite{DEL:H}.

We will focus on two specific examples with $k=24$. First we will consider
degeneration of the $\SU(2)$-bundle to that of 24 ``point-like
instantons''. That is, this bundle becomes everywhere flat except at
24 points. We will denote the associated sheaf by
$\PIsh$.
The precise mathematical meaning of this will become
apparent. Second we will consider the tangent sheaf $\Scr{T}_Z$.

In order to compute the spectral curves we need to compute $F^1$ of
each of these sheaves. To do that we first consider a Fourier-Mukai
transformation at a generic elliptic fibre, $E$, of $Z$.

First consider the point-like instanton sheaf
$\PIsh$. Clearly it restricts as
\begin{equation}
  \PIsh|_E \cong \O_E\oplus\O_E,
\end{equation}
so long as we are not at one of the 24 points where the instantons are
located.

Now for the tangent sheaf case we have the exact sequence
\begin{equation}
  0 \to \Scr{T}_E \to \Scr{T}_Z|_E \to \Scr{N}_{E/Z} \to 0,
		\label{eq:esT}
\end{equation}
where $\Scr{N}_{E/Z}$ is the normal sheaf of $E\subset Z$. Now since
$\Scr{T}_E\cong\Scr{N}_{E/Z}\cong\O_E$, we see that $\Scr{T}_Z|_E$ is
an extension of $\O_E$ by $\O_E$. In particular {\em on a generic
fibre, the tangent sheaf is $S$-equivalent to the point-like instanton
sheaf.}

To a string theorist it should come as a nasty surprise that the tangent
sheaf and the point-like instanton sheaf should be so similar. There
is certainly no hope that these two theories could somehow be dual to
each other. The heterotic string on the tangent bundle of a smooth K3
surface should be a very well-behaved theory with no extra massless
particles. The point-like instantons, on the other hand, are known to
leave an unbroken $E_8$ gauge symmetry and generate massless tensor
particles in six dimensions \cite{SW:6d,MV:F}.

One might hope that the full form of the spectral curve will get us
out of this difficulty but we will see that this is not the case.
Recall that we saw earlier that $F^1(\O_Z)$ produced a sheaf in $Z$
supported along $\sigma_0$.
Thus in each case we see that the generic elliptic fibres produce a
double copy (in
a way to be made precise) of $\sigma_0$ in the spectral curve. Thus,
from (\ref{eq:Ccl}) the non-generic contribution must be in the class
$24[E]$. The only divisor of $Z$ in this class is 24 copies of $E$. In
both cases therefore $C_S$ is a reducible curve containing $\sigma_0$
doubly and 24 fibres. The only question is which 24 fibres.

In the case of $\PIsh$ the answer is clear --- we use the
24 elliptic fibres containing the 24 instantons. In the case of
$\Scr{T}_Z$ for a generic K3 surface the answer is also obvious --- we
use the 24 I$_1$ fibres where the smooth elliptic fibres
degenerate.\footnote{This was shown explicitly using toric geometry in
\cite{BCGJL:F}.} These are the only locations where the above arguments
fail.

In general we see that the spectral curve for $\PIsh$
will be different than that for the tangent sheaf. Generically
$F^1(\PIsh)$ contains 24 smooth fibres whereas the 24
fibres in $\Scr{T}_Z$ are singular. We can remove this discrepancy however by
specializing to the case of $\PIsh$ where all 24 points
are located on singular fibres of $\pi:Z\to B$. Now  the spectral
curves for $\PIsh$ and $\Scr{T}_Z$ look identical!

We thus obtain
\begin{prop}
  Let $X_T$ be the \CY\ threefold which represents the F-theory dual of
  the $E_8\times E_8$ heterotic string compactified on the tangent
  bundle of a K3 surface. Let $X_I$ be the \CY\ threefold which
  represents the F-theory dual of the same heterotic string theory
  compactified on the same K3 surface except with one point-like
  instanton located at a point on each of the 24 I$_1$ fibres of the
  elliptic fibration $\pi:Z\to B$. Then $X_T$ is isomorphic to $X_I$.
\end{prop}

\begin{figure}
\begin{center}
\setlength{\unitlength}{0.00058300in}%
\begin{picture}(6285,3495)(1066,-4573)
\thinlines
\put(1366,-3361){\line( 1, 0){5835}}
\put(2101,-1261){\line( 0,-1){3300}}
\put(3301,-1261){\line( 0,-1){3300}}
\put(4501,-1261){\line( 0,-1){3300}}
\put(5701,-1261){\line( 0,-1){3300}}
\put(1381,-1561){
}
\put(1951,-3061){
}
\put(2807,-3818){
}
\put(1951,-3061){
}
\put(1726,-3436){
}
\put(5176,-3061){
}
\put(7276,-1636){\makebox(0,0)[lb]{\smash{$C_*$}}}
\put(2101,-1186){\makebox(0,0)[lb]{\smash{I$_1$}}}
\put(3301,-1186){\makebox(0,0)[lb]{\smash{I$_1$}}}
\put(4501,-1186){\makebox(0,0)[lb]{\smash{I$_1$}}}
\put(5701,-1186){\makebox(0,0)[lb]{\smash{I$_1$}}}
\put(6751,-3136){\makebox(0,0)[lb]{\smash{I$_1$}}}
\put(7351,-3511){\makebox(0,0)[lb]{\smash{$C_0$}}}
\put(950,-3511){\makebox(0,0)[lb]{\smash{II$^*$}}}
\end{picture}
\end{center}
  \caption{The discriminant for the tangent bundle and point-like instantons.}
\label{fig:D2}
\end{figure}
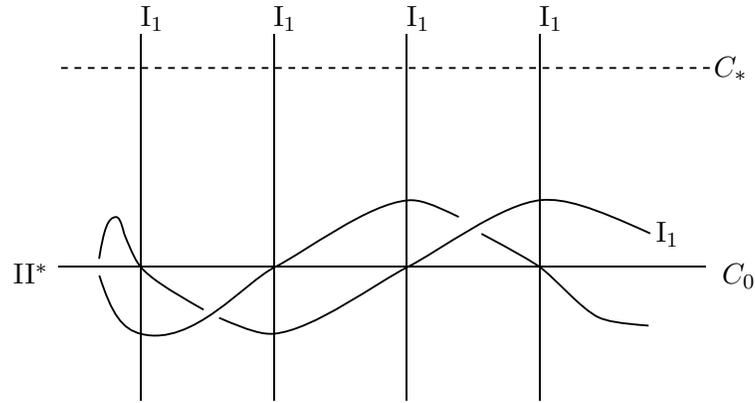

We draw the form of the discriminant locus $\Delta_1\subset\Sigma$ for
$X_1$ for this case in figure \ref{fig:D2}. This is also discussed in
\cite{me:hyp}.

Applying the usual rules of F-theory to figure \ref{fig:D2} we see
that we have a curve of type II$^*$ fibres along $C_0$ which implies
an $E_8$ gauge symmetry. We also have 24 I$_1$-II$^*$ collisions each
of which must be blown up to resolve $X$. This implies 24 massless
tensors. The result of an $E_8$ gauge symmetry and 24 massless tensors
is exactly what we would expect for the 24 point-like instantons. It
is not at all what we want for the tangent bundle. Clearly we have to
think a little more carefully about how the rules for F-theory should
be applied and, more importantly, what precisely should distinguish
the tangent bundle from the point-like instanton case.


\subsection{Ramond-Ramond moduli and the degenerate Jacobian}
\label{ss:RRJ}

The only thing in the type IIA language on $X$ that remains that can
possibly distinguish the two cases are the RR moduli. We know that
these parameterize the exact form of the spectral sheaf supported on
the spectral curve $C_S$. If $C_S$ were smooth, we would expect the RR
moduli to live on the torus representing the Jacobian of $C_S$. In
order for duality to work, the same thing should be true when $C_S$
degenerates except that this time the Jacobian will generally not be a
torus. As we have seen before, the genus of the spectral curve $C_S$
is $2k-3=45$. So in the smooth case the Jacobian is a 45 (complex)
dimensional torus, while in the singular case we expect the Jacobian
to be more complicated, but still 45 dimensional.

We thus expect $\Scr{T}_Z$ and $\PIsh$ to be distinguished
by the fact that their transforms under $F^1$ will differ as sheaves
even though the spectral curve, the support of this sheaf, may look
the same in both cases.

We begin by computing $F^1(\Scr{T}_Z)$. This was also discussed in
section 6.2 of \cite{FMW:Ell}. Let us analyze the Fourier-Mukai
transform acting on a single generic elliptic fibre,
$E$, of $X$ again --- this
time with a little more care.
As mentioned earlier, since the
restriction $\Scr{T}_Z|_E$ is an extension of $\O_E$ by $\O_E$ we expect
this to map, under $F^1$, to ``twice the origin'' of $\hat
E$. To avoid cluttering notation we will omit the restriction to $E$ for a
while. Restricting our earlier definition of the Fourier-Mukai
transformation to just a single elliptic fibre we have
\begin{equation}
  F^1(\Scr{T}_Z) = R^1\hat p_*(\pbun\otimes p^*\Scr{T}_Z).  \label{eq:FMx}
\end{equation}
where now $p:E\times\hat E\to E$ and $\hat p:E\times\hat E\to\hat
E$. This is a sheaf over $\hat E$ which has a trivial stalk except at
the origin. Let us further restrict this bundle to the origin, $\hat
0$.
Now
\begin{equation}
\begin{split}
  F^1(\Scr{T}_Z)|_{\hat 0} &= \Bigl(R^1\hat p_*(\pbun\otimes
  p^*\Scr{T}_Z)\Bigr)|_{\hat 0} \\
  &= H^1(E,\pbun|_{E\times\hat 0}\otimes\Scr{T}_Z)\\
  &= H^1(E,\Scr{T}_Z|_E),
\end{split}
\end{equation}
where we have used the ``base change'' theorem of \cite{Hartshorne:}
and the definition of the Poincar\'e bundle we stated earlier.

Now the exact sequence (\ref{eq:esT}) gives
\begin{multline}
  0 \to H^0(E,\Scr{T}_E) \to H^0(E,\Scr{T}_Z|_E) \to
  H^0(E,\Scr{N}_{E/Z}) \labto r H^1(E,\Scr{T}_E) \to\\
  H^1(E,\Scr{T}_Z|_E)
  \to  H^1(E,\Scr{N}_{E/Z}) \to 0.
\end{multline}
Since $\Scr{T}_E\cong\Scr{N}_{E/Z}\cong\O_E$ we see that
\begin{equation}
  \dim H^1(E,\Scr{T}_Z|_E) = 2-\rank(r).
\end{equation}
It is also true that $\Hom(H^0(E,\Scr{N}_{E/Z}),H^1(E,\Scr{T}_E))
=\Ext(\Scr{N}_{E/Z},\Scr{T}_E)$
and so $\rank(r)=0$ if (\ref{eq:esT}) splits and $\rank(r)=1$
otherwise.

The map $r$ is given by the Kodaira-Spencer theory of deformations. It
shows how for a family of elliptic curves $\pi:Z\to B$, the tangent
directions in the base $\Scr{N}_{E/Z}\cong\Scr{T}_B$ map into
deformations of $E$. Generically, our elliptic K3 surface will not
have constant fibre and so this map is not zero. Thus $\rank(r)=1$ and
$\dim H^1(E,\Scr{T}_Z|_E)=1$ for a generic $E$.

We have thus concluded that $F^1(\Scr{T}_Z)$ restricted to $\hat E$ is
a rather peculiar sheaf. It is of ``length'' two --- that is it has
two complex degrees of freedom, since $\Scr{T}_Z$ is rank two, and yet
it is supported only at a single point $\hat 0\in\hat E$ whereupon it
restricts to a sheaf of rank one. How can this be?

The answer is that it is the skyscraper sheaf on a ``fat'' point. Let
$\Scr{I}^2_{\hat 0}$ be the sheaf of functions which vanish at $\hat
0$ and whose derivative also vanishes at $\hat 0$. The sheaf we
require is $\O_{\hat E}/\Scr{I}^2_{\hat 0}$.

The upshot of all this is that we may state more clearly what we
meant by saying that the spectral curve $C_S$ of $F^1(\Scr{T}_Z)$
contains $\sigma_0$ ``doubly''. $C_S$ really contains a ``fat'' copy
of the line $\sigma_0$. The sheaf $F^1(\Scr{T}_Z)$ may be viewed as a
sheaf supported on this fat $\sigma_0$ plus 24 I$_1$ fibres. Over this
support the sheaf can be thought of as having rank one.

Let us contrast this to the transform of the point-like instantons.
All of the
analysis above may be repeated except that this time the sequence
defining $\PIsh$ {\em is\/} generically split and so
$r=0$. This means that \newline $\dim H^1(E,\PIsh|_E)=2$ and so
over a generic $\hat E$, $F^1(\PIsh)$ looks like
two copies of the skyscraper sheaf, $\O_{\hat 0}\oplus\O_{\hat 0}$.

We see that the sheaf $F^1(\PIsh)$ is a sheaf whose
support contains the reduced (non-fat) $\sigma_0$ but it has rank two
over this curve.

The two sheaves $F^1(\Scr{T}_Z)$ and $F^1(\PIsh)$ are
therefore qualitatively different objects even though na\"\i vely
their support looks the same. Fixing this spectral curve, each of
these sheaves may be varied in a 45 parameter family. This is what we
would expect for the (complex) dimension of the RR moduli
space. However, since these two families are quite different we must
have at least two {\em components\/} to the Jacobian in which the RR
moduli live. We thus resolve our problem with

\begin{prop}
  When $X\cong X_1\cong X_T$, the moduli space of RR fields has at
  least two components. The two points corresponding to the tangent
  bundle and the point-like instantons lie in different
  components.
\end{prop}

In particular, the moduli space of RR fields need not be a simple
torus when $X$ is singular. Its torus part is, in general, only 24
complex dimensional: it is the Jacobian of the reduced part of the
spectral curve $C_S$, i.e. the product of the $24$ elliptic fibers.
(In the case we are actually considering, all $24$ fibers happen to
be singular, so this $24$ dimensional part of the Jacobian is no
longer compact, but rather the product of $24$ copies of $\C^*$.)
The remaining $21$ dimensions come from the non-reduced structure
along $\sigma_0$. It is this part which is reducible, with one
component corresponding to deformations of the tangent bundle and
another corresponding to deformations of the point-like instantons.
We will see a more detailed description of these $21$ parameters
for deforming the point-like instanton in the next subsection.


\subsection{Obstructed extremal transitions}  \label{ss:ext}

Having sorted out why the tangent bundle is not the same thing as 24
point-like instantons as far as string duality is concerned, we need
to tidy up the rules of F-theory. The heterotic string on a tangent
bundle should not have massless tensors or totally unbroken gauge
symmetry even though figure \ref{fig:D2} suggests it should.

Let us consider the general picture of an extremal transition. Begin
with a smooth three-dimensional \CY\ manifold $Y$. Deform the complex
structure of
$Y$ to produce a singular variety $Y^\sharp$. In some cases
$Y^\sharp$ can be resolved by blowing up to produce a smooth \CY\
threefold $Y'$ which is not topologically (or birationally) equivalent
to $Y$. It is not hard to see that $h^{2,1}(Y')<h^{2,1}(Y)$ and
$h^{1,1}(Y')>h^{1,1}(Y)$.

Now when we put such an extremal transition in the context of type IIA
compactifications we need to worry about the RR fields. Since
$h^{2,1}(Y')<h^{2,1}(Y)$ it must be that the type IIA string
compactified on $Y'$ must have fewer RR degrees of freedom than
$Y$. Thus, in order to pass through the extremal transition some of
the RR parameters {\em must\/} be tuned to a fixed value --- which we
call zero.

Now we wish to claim that an enhancement of gauge symmetry corresponds
to an extremal transition.
To see this we think again in the 4-dimensional type IIA
picture corresponding to the heterotic string compactified on $Z$
times a 2-torus. When the structure group of the bundle decreases, the
gauge symmetry is enhanced and we may switch on more vector moduli
(which break the enhanced gauge symmetry back down to its Cartan
subgroup). This corresponds to blowing up the fibre in the elliptic
fibration of $X$ \cite{MV:F} (see also \cite{me:lK3}).

This potentially explains why the tangent sheaf doesn't give rise to
the $E_8$ gauge symmetry. The blow-up in the fibre which raised the
gauge symmetry from $E_7$ to $E_8$ completes an extremal transition
which kills some of
the RR fields which are not zero for the tangent sheaf. We will now
try to justify this claim by arguing that these RR fields {\em
are\/} zero for the point-like instantons.

First we wish to make a change to our concept of the point-like
instanton. Until now we have really considered it to be an object that
lives on the boundary of the moduli space of $\SU(2)$-bundles. While
this is true, the point-like instanton is really no more an
$\SU(2)$-object than an $\SU(3)$-object or an object tied to any
nontrivial group. Indeed, as the holonomy of a point-like instanton is
trivial it would be more natural to associate it to a trivial
structure group.

We will claim the following:
\begin{prop}
  A point-like instanton is the ideal sheaf of a point, $\Scr{I}_z$.
\end{prop}
In particular we claim that
\begin{equation}
  \PIsh \cong
  \O_Z\oplus\Scr{I}_{z_1,z_2,\ldots,z_{24}},  \label{eq:idsh}
\end{equation}
where $\Scr{I}_{z_1,z_2,\ldots,z_{24}}$ is the ideal sheaf of
functions vanishing at $z_1,z_2,\ldots,z_{24}$ which are the locations
of the point-like instantons. We no longer insist that these points
lie in the I$_1$ fibres.

Clearly this claim is perfectly reasonably away from the
instantons since both sides of (\ref{eq:idsh}) are then equal to
$\O_Z\oplus\O_Z$. It follows that the spectral curve is a copy of
$\sigma_0$, over which the Fourier-Mukai transformed sheaf has rank 2,
and the 24 elliptic fibres containing the $z_i$'s. The above argument
implies that they live in the same component of the RR moduli space
for this fixed spectral curve.

Let us now consider how we might parameterize this 45 dimensional
component of the RR moduli space. Clearly 24 directions are given by
moving the locations of the $z_i$ up and down the fibres. Locally near
each instanton we may view the sheaf as
$\O_Z\oplus\Scr{I}_{z_i}$. However there is no reason why we need to
globally insist that the sheaf decomposes as a sum over the whole of
$Z$. At each of the 24 points $z_i$ choose a direction
$\C\subset\C^2$ (i.e., a point on $\P^1$) which specifies how we
locally decompose the rank 2 sheaf into $\O_Z\oplus\Scr{I}_{z_i}$. By
an $SL(2,\C)$ symmetry action on this $\C^2$ fibre, this gives us
$24-3=21$ more parameters.

These numbers work perfectly for comparison with the $E_8$
transition. When the $E_8$ appears we do the
blow-up in the fibre, $h^{2,1}$ decreases by 21 and we lose 21 complex
parameters
from the RR moduli. That is, in order to switch on the $E_8$
transition, the sheaf must {\em globally\/} decompose as
$\O_Z\oplus\Scr{I}_{z_i}$ in agreement with (\ref{eq:idsh}).

In many ways this gives the ideal sheaf the interpretation of
an ``$\SU(1)$-bundle''.
Consider the moduli space of the sheaf, $\shF(N)$, corresponding to an
$\SU(N)$ bundle. There is a locus within this moduli space where the
sheaf decomposes $\O_Z\oplus\shF(N-1)$. This corresponds to
perturbative enhanced gauge symmetry for the heterotic string.
For example an $\SU(3)$-bundle will leave an $E_6$ gauge symmetry
unbroken but if we tune the bundle moduli we may decrease its
structure group to $\SU(2)$ and obtain an unbroken $E_7$ gauge symmetry.
Given
this behaviour, it is natural to interpret the ideal
sheaf, $\Scr{I}_{z_1,z_2,\ldots,z_{24}}$, as a degenerate $\SU(1)$ bundle with
$c_2=24$. The only feature particular to point-like instantons is that the
extremal transition is more complicated and leads to massless tensors
in addition to more gauge symmetry.

We hope the reader agrees that this is reasonable evidence to assert
that the point-like instanton case really is the same thing as the
ideal sheaf and that this is where we identify the RR-fields as having
value ``zero'' allowing the extremal transition.

Having accepted this, it follows that the tangent bundle does {\em
not\/} have the correctly-tuned RR fields to allow the extremal
transitions giving rise to the massless tensor or the $E_8$ gauge
symmetry. Indeed it is quite apparent why this is so. One cannot
possibly try to decompose the sheaf $F^1(\Scr{T}_Z)$ as it has this
strange part looking like a rank one sheaf over a ``fat'' non-reduced
curve. The two degrees of freedom we expect from this part of the
sheaf are inextricably wound together. This can be contrasted to the
point-like instantons where we have the sum of two copies of the
structure sheaf over the same curve.


\section{The tangent sheaf of an orbifold K3}  \label{s:orb}

As an application of the technology we have pursued above let us
consider the six-dimensional physics of the heterotic string
compactified on the
tangent bundle of a K3 surface when we go to an orbifold limit.

Given that we know explicitly how to identify the K3 moduli in the
F-theory language \cite{FMW:F,AM:po,me:hyp}, it is straight-forward to
identify the exact form of the required \CY\ threefold, $X$, or more
precisely its stable degeneration, $X_1\cup X_2$, required to make the
K3 surface have
a particular complex structure compatible with the elliptic
fibration. We will consider the case where the complex structure puts
an orbifold point in $Z$.

Let $s$ and $t$ be affine coordinates for a patch on $\Sigma\cong\HS
n$ (where $n=-12$ to obtain the correct $c_2$) as in \cite{AM:po} and
let our elliptic fibration,
$\pi_F^1:X_1\to\Sigma$  be in standard Weierstrass form
\begin{equation}
  y^2 = x^3 + a(s,t)\,x+b(s,t).    \label{eq:We1}
\end{equation}
This has a discriminant equal to $4a^3+27b^2$.
Restricting this fibration to $C_*$ we obtain the Weierstrass form for
the K3 surface $Z$. We may let $t$ be an affine coordinate on $C_*$.
It is not difficult to show
\begin{theorem}
  Consider the heterotic string compactified on the tangent sheaf of the
  K3 surface $Z$ which has Weierstrass form
\begin{equation}
  y^2 = x^3 + \alpha(t)\,x + \beta(t).
\end{equation}
  This is dual to F-theory compactified on the \CY\ threefold $X$
  which, in the stable degeneration, gives
  $X_1$ in the form (\ref{eq:We1}) with
\begin{equation}
\begin{split}
  a(s,t) &= \alpha(t)s^4\\
  b(s,t) &= \beta(t)s^6 + \Bigl(4\alpha(t)^3 + 27\beta(t)^2\Bigr)s^5.
\end{split}
\end{equation}
\end{theorem}

This may be proven by imposing the condition that the fibration of $X_1$
reduces to that of $Z$ when restricted to $C_*$ (where $s=\infty$) and
by analyzing the form of the discriminant to obtain the
required factorization shown in figure \ref{fig:D2}.

\begin{figure}
\begin{center}
\setlength{\unitlength}{0.00050000in}%
\begin{picture}(6075,3195)(1501,-3373)
\thinlines
\put(1801,-2761){\line( 1, 0){5700}}
\put(5101,-361){\line( 0,-1){3000}}
\put(1801,-661){
}
\put(3001,-361){\line( 0,-1){3000}}
\put(2026,-1561){
}
\put(3901,-1988){
}
\put(7576,-2836){\makebox(0,0)[lb]{\smash{$C_0$}}}
\put(7576,-736){\makebox(0,0)[lb]{\smash{$C_*$}}}
\put(1501,-2836){\makebox(0,0)[lb]{\smash{II$^*$}}}
\put(1801,-1636){\makebox(0,0)[lb]{\smash{I$_1$}}}
\put(3001,-286){\makebox(0,0)[lb]{\smash{I$_1$}}}
\put(5101,-286){\makebox(0,0)[lb]{\smash{I$_2$}}}
\end{picture}
\end{center}
  \caption{The discriminant for the tangent sheaf on a $\C^2/\Z_2$
  singularity.}
\label{fig:DZ2}
\end{figure}
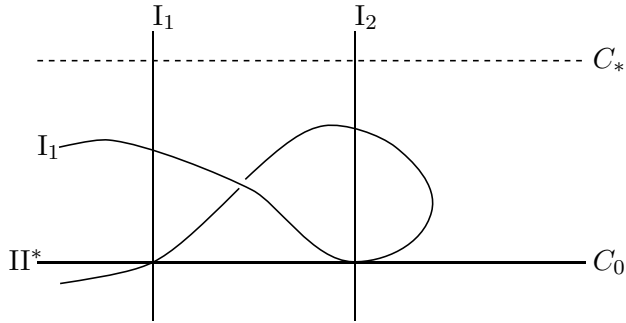

We show the discriminant locus for $X_1$ in figure \ref{fig:DZ2} where
$Z$ acquires a $\C^2/\Z_2$ singularity.  Locally this may be modeled
by putting $\alpha=-3$ and $\beta=2+t^2$.  The K3 surface has a
$\C^2/\Z_2$ singularity because the I$_2$ fibre intersects $C_*$
which is where $Z$ sits in the elliptic fibration.

Applying the usual rules of F-theory to figure \ref{fig:DZ2} would
require us to blow-up the collision between the curves of II$^*$ and
I$_2$ fibres twice to resolve the threefold. This would imply a local
contribution of
\begin{itemize}
  \item A nonperturbative $\SU(2)$ gauge symmetry in addition to the
  full unbroken perturbative $E_8$ gauge symmetry.
  \item Two massless tensors.
  \item Four hypermultiplets in the $\mathbf{2}$ representation of the
  above $\SU(2)$.
\end{itemize}

Now this would seem to be somewhat excessive! The problem is of course
that we need to worry about the RR moduli again and the fact that they
may block various extremal transitions giving rise to the above spectrum. It
seems reasonable to assume that the above massless spectrum actually
corresponds to the point-like instanton case. That is, two point-like
instantons coalesce at the quotient singularity. This was also
asserted in \cite{AM:po} (without worrying about RR moduli).

Now in many ways figure \ref{fig:DZ2} is the same thing as figure
\ref{fig:D2} except that two of the vertical I$_1$ lines have
coalesced. Therefore much of the discussion in section \ref{ss:RRJ}
applies. This can be used to argue that for the tangent sheaf case we
do {\em not\/} have any massless tensors and the $E_8$ in not fully
restored --- we only have an $E_7$ part of it as for any generic
$\SU(2)$-bundle.

An interesting question is whether we actually have the
nonperturbative $\SU(2)$ gauge symmetry and its associated
hypermultiplets. We now argue that we do.

The problem we need to address is how to identify which RR moduli are
lost when we go through the extremal transition associated to this
$\SU(2)$ gauge symmetry enhancement. As mentioned earlier, in order to
see the extremal transition we need to go to four-dimensional physics
given by the type IIA string compactified on $X$ (where the fibres of
the F-theory elliptic fibration can be given nonzero size).

If we really have an $\SU(2)$ gauge symmetry in the four dimensional
theory then we should be able to break it to $\GU(1)$ by switching on
a vector multiplet. In terms of $X_1$, this is the deformation of the
K\"ahler form which blows up all the I$_2$ fibres along the vertical
line in figure \ref{fig:DZ2}. As this line passes through $C_*$ it
corresponds to blowing up to produce an algebraic $\P^1\subset Z$.

Recall that in \cite{me:hyp} some of the RR modes were identified with
3 cycles which intersected $Z$ along transcendental 2-cycles. As we
have produced an algebraic 2-cycle in $Z$ in this extremal transition
we must have lost a transcendental 2-cycle. This lost 2-cycle is naturally
the one which lay in the 3-cycle corresponding to the lost RR mode.

The analysis in \cite{me:hyp} argued that the RR fields
associated to 3-cycles which intersect $Z$ in 2-cycles are interpreted
as B-fields in the heterotic language. That is, {\em the RR mode which
potentially blocks the appearance of the nonperturbative $\SU(2)$
gauge symmetry is a $B$-field on the K3 surface.} (We expect the masslessness
of the four $\mathbf{2}$'s of $\SU(2)$ to be blocked in the same way
since there are no other moduli which could further affect these modes.)

The conclusion is therefore that in order to get enhanced gauge symmetry
for the tangent sheaf of an orbifold, the corresponding component of the
$B$-field must be set to zero. This is highly reminiscent of the type
IIA string on a K3 surface as described in \cite{me:enhg,W:dyn2} where
it was also argued that the $B$-field must also be set to zero to see the
enhanced gauge symmetry from an orbifold point.

It is not difficult to generalize this discussion to that of a general
A-D-E singularity. We find
\begin{prop}
  If the $E_8\times E_8$ heterotic string is compactified on the
  tangent sheaf of a K3 surface with an A-D-E singularity then one
  acquires the corresponding A-D-E gauge group nonperturbatively if
  and only if the $B$-field associated to the shrunken 2-cycle(s) is
  set to zero.
\end{prop}

For the cyclic quotient singularity we expect an $\SU(N)$ gauge
symmetry to come with $2N$ hypermultiplets in the fundamental
representation. This is implied by the F-theory geometry and and can
also be seen to correctly cancel
anomalies. Similarly F-theory suggests $2N-8$ hypermultiplets in the
vector representation of $\SO(2N)$ and no hypermultiplets in the $E_6$,
$E_7$, and $E_8$ case.

It is also perhaps worth mentioning a local description of the tangent
sheaf of a quotient singularity in terms of previously analyzed
point-like instantons. In \cite{AM:frac} methods were described which
analyze a point-like instanton with a $\Z_2$ holonomy which broke
$E_8$ to $(E_7\times\SU(2))/\Z_2$.  Such an instanton is forced to sit
on a $\C^2/\Z_2$ quotient singularity (or worse).  By using methods
similar to section 4.1 of \cite{AM:frac} one may show that this
instanton has $c_2=\ff12$. Figure \ref{fig:DZ2} is precisely the
F-theory picture for what happens when one allows such an instanton to
coalesce with a normal free point-like instanton in the $E_8\times
E_8$ heterotic string. Therefore up to changes in the RR-moduli, the
tangent sheaf of $\C^2/\Z_2$ looks locally like a sum of a ``half''
instanton with a full instanton. It therefore has a local charge of
$c_2=\ff32$. This is consistent with the construction of a K3 surface
as a quotient $T^4/\Z_2$. This has 16 orbifold points of the above
form and the Euler characteristic is 24 --- so each orbifold point
``contributes'' $\ff32$ to the Euler characteristic.


\section{T-Duality}	\label{s:T}

Finally we will study some global issues about the moduli
space of the heterotic string compactified on the tangent bundle of a
K3 surface. We will discover that the mathematical abstraction of
derived categories is pretty well forced upon us if we are to make
sense of conventional ideas such as T-duality.

Consider first the \nlsm\ of a superstring with a K3 target
space. When the metric on this K3 surface is Ricci-flat, we
have a conformal field theory which has $N=(4,4)$ supersymmetry. It
was argued in \cite{AGG:N=4,Hull:N=4,BS:N=4} that the moduli space of
such conformal field theories receives no quantum corrections. It was
then deduced in \cite{Sei:K3,AM:K3p} that the moduli space must be
exactly of the form
\begin{equation}
  \cM\cong\GO(\Gamma_{4,20})\backslash\GO(4,20)/(\GO(4)\times\GO(20)),
		\label{eq:K3m}
\end{equation}
where $\Gamma_{4,20}$ is the even unimodular lattice of signature
$(4,20)$ and $\GO(\Gamma_{4,20})$ is its discrete isometry group
(often called, more loosely, $\GO(4,20;\Z)$).

The best way to view the action of $\GO(\Gamma_{4,20})$ is as
follows. Consider $H^*(Z,\Z)=H^0(Z,\Z)\oplus H^2(Z,\Z)\oplus
H^4(Z,\Z)$ and let $w\in H^*(Z,\Z)\cong\Gamma_{4,20}$ be a primitive
null vector which denotes the
generator of $H^4(Z,\Z)$. Given a point in the moduli space and a
choice of $w\in\Gamma_{4,20}$, we may define a
Ricci-flat metric on $Z$ and a choice of $B$-field
\cite{AM:K3p,me:lK3}. The modular group $\GO(\Gamma_{4,20})$ acts
transitively on the set of possible choices for $w$ within the lattice
$\Gamma_{4,20}$ and so generates the possible target space
interpretations of a single point in the moduli space. These
identifications include such notions as mirror symmetry and
$R\leftrightarrow1/R$ symmetries.

The $N=(4,4)$ conformal field theory has no preference for which
direction in $H^*(Z,\Z)$ is $H^4(Z,\Z)$ --- this is what the modular
group $\GO(\Gamma_{4,20})$ expresses. In order to impose a geometric
interpretation on the theory we are forced to choose a $w$, breaking
this modular group.

Any string theory compactified in a conventional way should have the
moduli space of corresponding conformal field theories as part of its
moduli space --- supplemented by more parameters such as the dilaton,
RR fields etc. The conformal field theory moduli space may receive
corrections which are quantum with respect to the string coupling (as
distinct from the $\sigma$-model coupling).
Given enough supersymmetry, such quantum corrections may vanish
leaving the conformal field theory moduli space intact.

An example of this is the type IIA string compactified on a K3
surface. The well-known duality of this to the heterotic string on a
4-torus requires that the moduli space (\ref{eq:K3m}) be exact.

We can also expect to rid ourselves of such quantum corrections if the
dilaton somehow lives in another sector of the theory. This happens
when the heterotic string is compactified on a K3 surface --- the
dilaton lives in a tensor multiplet whereas the conformal field moduli
space is composed of hypermultiplet moduli. When we ``embed the spin
connection in the gauge group'' for the heterotic string, we obtain an
$N=(4,4)$ underlying conformal field theory and so the space
(\ref{eq:K3m}) {\em must\/} be an exact subspace of the moduli space
of heterotic strings on a K3 surface.

Note that when we deform the bundle of the heterotic string away from
the tangent bundle, we generically break the world-sheet supersymmetry
to $N=(4,0)$ and then the moduli space may well pick up quantum
corrections in the \nlsm. This was discussed in \cite{me:hyp}.

The question we want to address is, can we extend the
T-duality for conformal field theories to T-duality for heterotic
strings?
The answer is strikingly clear. The Mukai vector lives in $H^*(Z,\Z)$,
We must therefore let
the T-duality group act on this too.

It is worth pointing out that Mukai's work of \cite{Muk:bun} precedes
the notion of mirror symmetry and that the appearance of $H^*(Z,\Z)$ in
both contexts appeared to be somewhat unrelated at first. One
connection between these objects has already been pointed out by
Morrison \cite{Mor:SYZ} based on the work \cite{SYZ:mir}. Here we have
another relation --- perhaps more unavoidable because of the way that
the heterotic string combines vector bundles with conformal
field theories.

As explained in section \ref{ss:der}, the group
$\GO(\Gamma_{4,20})$ does not act in a natural way on the space of vector
bundles or even the space of sheaves. Instead we are forced to use
the derived category of coherent sheaves. We can then make the
following
\begin{prop}
  Let $Z_1$ be a K3 surface and
  consider the heterotic string compactified on the tangent bundle of
  $Z_1$. This is mapped by
  T-duality to an element, $W$, of the derived category of coherent sheaves
  over $Z_2$. $Z_2$ is related to $Z_1$ by the usual action of
  $\GO(\Gamma_{4,20})$ on the moduli space of conformal field
  theories. The Mukai vector of $W$ is obtained by the action of
  $\GO(\Gamma_{4,20})$ on
  the Mukai vector of the tangent sheaf of $Z_1$.
\end{prop}

It is worth pointing out that the way one might generate
$\GO(\Gamma_{4,20})$ in terms of conventional T-dualities on a  K3
surface and how one might generate it from actions on the Mukai vector
are remarkably similar. It was shown in \cite{AM:K3p} how
$\GO(\Gamma_{4,20})$ could be generated from three sets of elements:
\begin{enumerate}
  \item The classical diffeomorphisms of K3 generating \newline $\GO^+(H^2(Z,\Z))
	\cong\GO^+(\Gamma_{3,19})$.
  \item Mirror Symmetry.
  \item Shifts in the $B$-field by an element of $H^2(Z,\Z)$.
\end{enumerate}
It is not hard to see that the map $\mathbf{S}$ of section
\ref{ss:der} is similar to mirror symmetry\footnote{This similarity
can be a little confusing. Whether or not one really identifies it
with mirror symmetry comes down to a question of the actual definition
of mirror symmetry for a K3 surface.
In \cite{Mor:SYZ} an identification with mirror symmetry was made.
Having said that, $\mathbf{S}$ is not mirror
symmetry in the sense of mirror symmetry between families of algebraic
K3's as discussed in \cite{Dol:K3m} for example.}
and the map $\mathbf{T}$ is
the analogue of a shift by the class of the elliptic fibre.

We can now try to resolve an important question as to how, in general, the
heterotic string can be viewed. In particular what kind of data should
be specified in order to compactify it? The conventional point of view
was that a vector bundle over some \CY\ manifold was the correct
picture. It was seen in \cite{DGM:sh} that this is not sufficient to
fill out the moduli space in the case of compactification over \CY\
threefolds. Some phases of the moduli space correspond to
non-locally-free sheaves. In this case the sheaves were ``reflexive'',
i.e., they were locally free except over codimension 3 (that is, points).

One might therefore suspect that reflexive sheaves might be the right
choice. If one believes that string theory really has an algebraic
underpinning then one might also propose that coherent sheaves are the
natural choice.

Instead we propose that one must go even further and specify an element
of the derived category of coherent sheaves as the heterotic string
data. One must do this to get T-duality.

Let us spell out a little more clearly what we mean by this. Suppose
one takes a large smooth K3 surface with its tangent bundle. This
bundle is a locally-free sheaf. We now shrink the K3 surface to a size
well below the $\alpha'$ scale. We know that we may use T-duality to turn
such a K3 surface back into a large K3 surface. To do this however we
are required to reinterpret the way that we divided
$\Gamma_{4,20}\cong H^*(Z,\Z)$ into $H^0(Z,\Z)\oplus H^2(Z,\Z)\oplus
H^4(Z,\Z)$. For example, a good $R\leftrightarrow1/R$ symmetry is one
which exchanges the r\^ole of $H^0$ and $H^4$ \cite{me:lK3}. When we
do this however we are also forced to reinterpret the bundle data. A
vector bundle has Chern classes for example which live in $H^*(Z,\Z)$
and so these must be changed in accordance with T-duality. The natural
way to do this remapping is to use the Mukai vector and the derived category
$\Dr(Z)$.

Once we have accepted this claim for the tangent sheaf, we are
required to extend it to many other, if not all, possible sheaves on
which we may compactify the heterotic string. This is because we may
deform the tangent bundle into other $\SU(2)$-bundles and then we may
connect via extremal transitions to many other cases. It would appear
that the Mukai vector and $\Dr(Z)$ provides a good general setting for
analysis of the heterotic string.

We close with the observation that this is not the first time that
the derived category of coherent sheaves has appeared in string
theory. It was used by
Kontsevich in a conjectural
description of mirror symmetry \cite{Kon:mir}, which has recently been
proved for the one-dimensional case in
\cite{AP:Cmir}. While such abstract objects as derived categories are
not the kind of things a physicist would normally wish to consider, this
clearly deserves to be studied further!


\section*{Acknowledgements}

It is a pleasure to thank R.~Hain, D.~Morrison, T.~Pantev, R.~Plesser,
and D.~Reed for useful conversations.


\end{document}